\documentclass{article}

\PassOptionsToPackage{numbers, compress}{natbib}

\usepackage[table,xcdraw]{xcolor}

     \usepackage[final]{neurips_2022}

\usepackage[utf8]{inputenc} 
\usepackage[T1]{fontenc}    
\usepackage{hyperref}       
\usepackage{url}            
\usepackage{booktabs}       
\usepackage{amsfonts}       
\usepackage{nicefrac}       
\usepackage{microtype}      
\usepackage{xcolor}         
\usepackage{graphicx}
\usepackage{caption}
\usepackage{subcaption}
\usepackage{floatrow}
\newfloatcommand{capbtabbox}{table}[][\FBwidth]

\title{Knowledge distillation for fast and accurate DNA sequence correction}

\author{%
  Anastasiya Belyaeva\thanks{These authors contributed equally.} \\
  Google Research \\
  \And
  Joel Shor$^*$ \\
  Verily Life Sciences \\
  \And
  Daniel E. Cook \\
  Google Research \\
  \And
  Kishwar Shafin \\
  Google Research \\
  \And
  Daniel Liu \\
  Google Research \\
  \And
  Armin Töpfer \\ 
  Pacific Biosciences \\
  \And
  Aaron M. Wenger \\
  Pacific Biosciences \\
  \And
  William J. Rowell \\
  Pacific Biosciences \\
  \And
  Howard Yang \\
  Google Research\\
  \And
  Alexey Kolesnikov  \\
  Google Research \\
  \And
  Cory Y. McLean  \\
  Google Research \\
  \And
  Maria Nattestad \\
  Google Research \\
  \And
  Andrew Carroll  \\
  Google Research \\
  \And
  Pi-Chuan Chang  \\
  Google Research \\
}

\begin{document}

\maketitle

\begin{abstract}
Accurate genome sequencing can improve our understanding of biology and the genetic basis of disease.
The standard approach for generating DNA sequences from PacBio instruments relies on 
HMM-based models. Here, we introduce Distilled DeepConsensus - a distilled transformer–encoder model for sequence correction, which improves upon the HMM-based methods with runtime constraints in mind. \textbf{Distilled DeepConsensus is 1.3x faster and 1.5x smaller} than its larger counterpart while \textbf{improving the yield of high quality reads (Q30) over the HMM-based method by 1.69x (vs. 1.73x for larger model)}. With improved accuracy of genomic sequences, Distilled DeepConsensus improves downstream applications of genomic sequence analysis such as \textbf{reducing variant calling errors by 39\% (34\% for larger model)} and \textbf{improving genome assembly quality by 3.8\% (4.2\% for larger model)}. We show that the representations learned by Distilled DeepConsensus are similar between faster and slower models.

\end{abstract}

\section{Introduction}

Genome sequencing is widely used to advance our understanding of biological processes, the genetic basis of diseases, and is increasingly used for personalized medicine directly in the clinic~\cite{gorzynski2022ultrarapid}. Thus, obtaining accurate genome sequences is of paramount importance. 
Over the years, sequencing platforms have been developed by Illumina, Pacific Biosciences (PacBio), and Oxford Nanopore, that have had either short and accurate reads or long and noisy reads. Recently, PacBio developed a novel long read HiFi sequencing technology~\cite{wenger2019accurate} that has led to new state-of-the-art performance in variant calling~\cite{olson2022precisionfda} and the first telomere-to-telomere human assembly~\cite{nurk2022complete}. While highly accurate, some errors still exist and here we focus on whether we can use machine learning, and in particular transformers, to further improve the accuracy of the genomic sequences with on-device runtime constraints in mind.


PacBio HiFi sequencing works by circularizing DNA, allowing for multiple sequencing passes over the same molecule. This produces a set of noisy sequences/subreads of the same region. The subreads are then combined into a consensus sequence/read using a hidden Markov model (HMM)~\cite{wenger2019accurate}. 
In order to further improve upon the accuracy of the HMM-based method for correct DNA sequence generation, DeepConsensus - an encoder-only transformer model has been developed~\cite{baid2022deepconsensus}. However, the accuracy and yield improvements of DeepConsensus 
come with latency and model-size costs. 
For example, unlike HMMs, transformer models are too large to be run on sequencing platforms with resource constraints. As sequencing throughput increases, the problem compounds. 


In this work, we use Knowledge Distillation (KD)~\cite{hinton2015distilling} to improve the runtime of the transformer-based DeepConsensus model, using techniques similar to \cite{sanh2019distilbert, jiao2019tinybert, sun2020mobilebert}.
Our model is much smaller and faster than the original model, with slight performance degradation. Importantly, these runtime boosts are critical for enabling real time on-sequencing-device inference. 
In other domains of biology, as the amount of data increases, such as sequencing millions of single cells, and model size increases~\cite{avsec2021enformer}, the runtime considerations may become increasingly important. This work explores initial steps in this direction and our contributions are as follows:

\begin{enumerate}
\item 
We show that knowledge distillation can improve model accuracy, resulting in a faster model without a significant degradation in performance as compared to a large model.
\item We demonstrate relative performance parity between larger and smaller models on downstream tasks like variant calling and \textit{de novo} assembly generation.
\item We show that the representations learned by larger and smaller models are similar.
\end{enumerate}

\begin{figure}
     \centering
     \begin{subfigure}[b]{0.33\textwidth}
         \centering
         \includegraphics[width=\textwidth]{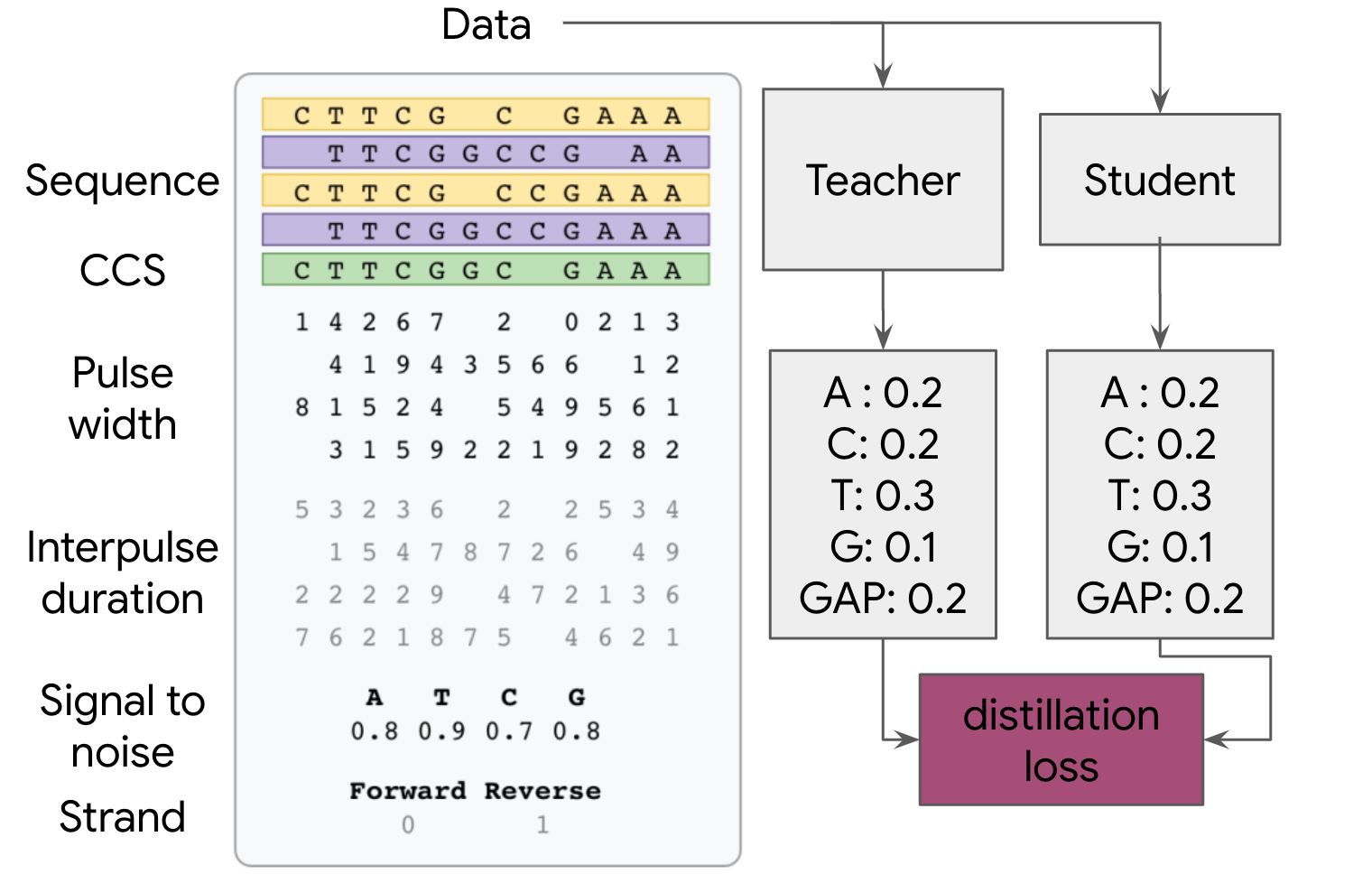}
         \caption{}
     \end{subfigure}
     \begin{subfigure}[b]{0.38\textwidth}
         \centering
         \includegraphics[width=\textwidth]{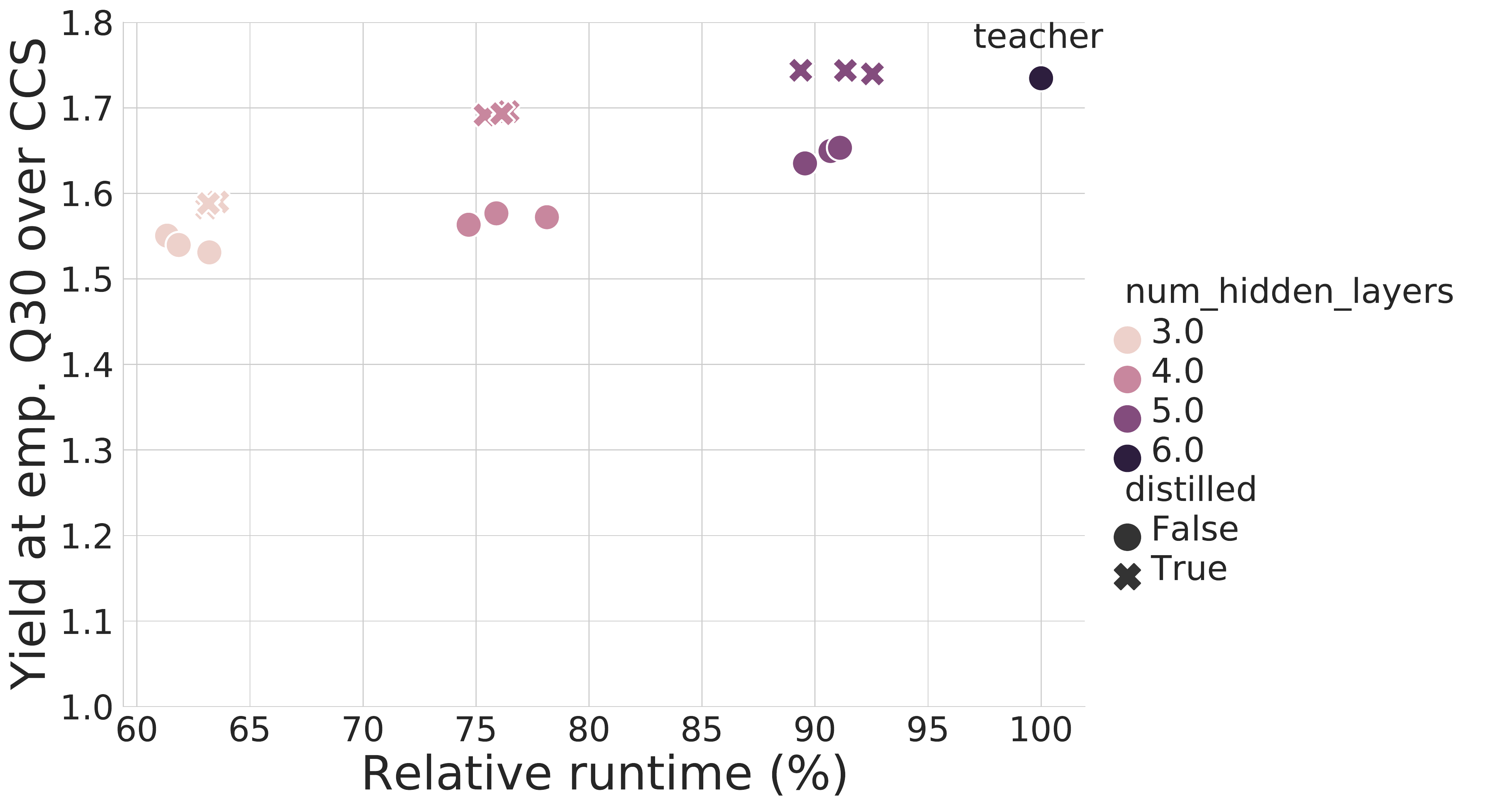}
         \caption{}
     \end{subfigure}
     \begin{subfigure}[b]{0.27\textwidth}
         \centering
          \includegraphics[width=\textwidth]{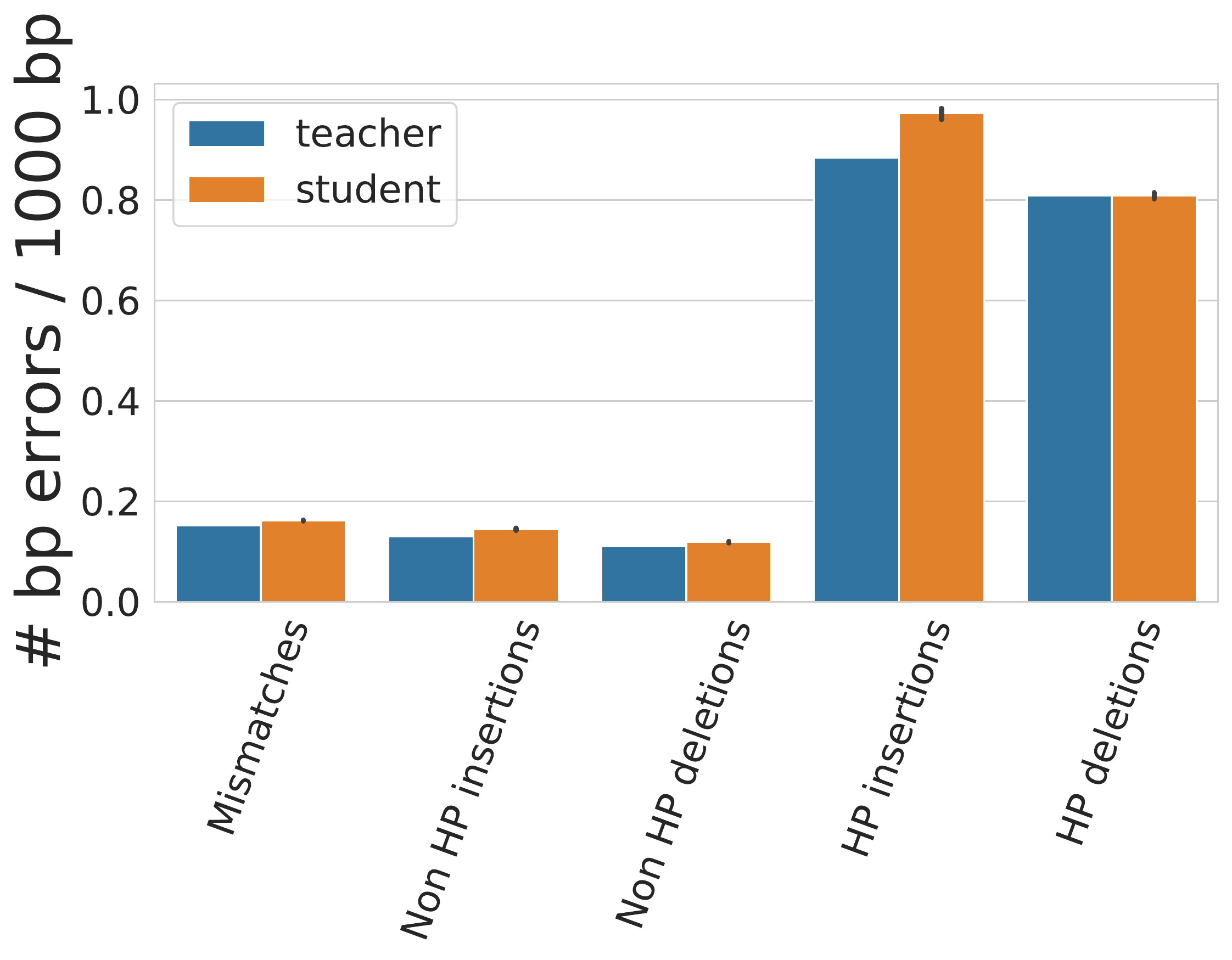}
         \caption{}
     \end{subfigure}
  \caption{(a) Overview of the Distilled DeepConsensus model. To train Distilled DeepConsensus, in addition to the alignment-based loss function we compute the distillation loss between teacher and student probabilities per position. (b) Performance versus runtime curve for Distilled DeepConsensus student models and undistilled 
  models. The performance is calculated relative to pbccs (HMM-based model). 
  (c) Error profiles of the teacher and distilled student models.
  }
  \label{fig:runtime_vs_accuracy}
\end{figure}

\section{Model}


DeepConsensus~\cite{baid2022deepconsensus} takes as input a tensor of length 100bp containing information about the subreads such as the noisy subread sequences, strand direction, pulse width, interpulse duration as well as general information such as signal-to-noise ratio of each nucleotide and the consensus sequence given by pbccs HMM-based algorithm (Fig.~\ref{fig:runtime_vs_accuracy}a). The goal is to predict the correct consensus sequence from the noisy subread sequences and the other auxiliary information. Since this problem is similar to sequence-to-sequence tasks, Transformers~\cite{vaswani2017attention} are leveraged for this problem.

Both the teacher and student models are encoder-only transformers. The teacher has six encoder stacks.
We vary the number of encoder stacks from 3 to 5 to obtain smaller (student) Distilled DeepConsensus models of different sizes. 
Each encoder stack is composed of a self-attention layer with hidden dimension of 280 and 2 attention heads and a feedforward network with a filter size of 2048. 
The output of the entire encoder block, which represents the contextual embedding of each base, is decoded with the feedforward layer followed by a softmax function, that for each base outputs the probability of A, C, T, G or a gap. 
We use this simpler decoder to save computational resources. 


For our labels, we leverage a high quality assembly of the human genome (T2T-CHM13), recently published by the Telomere-to-Telomere consortium~\cite{nurk2022complete}, as part of the effort to obtain the complete sequence of the human genome and use data (2 SMRT Cells) from a CHM13 sample with 24-kb insert sizes from~\cite{baid2022deepconsensus}.


We use an alignment-based loss function, which is based on differentiable dynamic programming~\cite{mensch2018differentiable}, $L(y, p)$ to train the teacher model as in~\cite{baid2022deepconsensus}, where $y = (y_1, ..., y_m)$ is the true sequence and $p = (p_1, ..., p_n)$ are the predicted probabilities with $p_i$ representing the probability over A, C, T, G or a gap classes.

In order to train smaller student models from the teacher, in addition to the alignment-based loss function, we use a typical knowledge distillation loss~\cite{hinton2015distilling}, to update the model weights of the student:

\begin{equation}
L_{total} = L(y, p^{student}) + \alpha L_{KD}(p^{student}, p^{teacher}),   
\end{equation}

where $p^{student}$ and $p^{teacher}$ are the probabilities output by the student and the teacher models, respectively, and $\alpha$ is determined using a hyper-parameter search. We explore both the Kullback–Leibler (KL) divergence and the mean squared error (computed per base and averaged across positions) for $L_{KD}$. In addition, we initialize the student layers from the teacher layers, since this was found to be beneficial in~\cite{sanh2019distilbert}.








\vspace{-1em}
\section{Results}

\vspace{-0.5em}
\subsection{Performance vs. runtime trade-off}

We measure the model performance by 
comparing the yield at empirical Q30 (number of bases in reads with Q30 or higher quality score, 
i.e.
identity of 0.999 or higher) produced by DeepConsensus models versus pbccs. This metric (yield at empirical Q30 over CCS) indicates how many more high quality bases 
we obtain with 
a particular model 
as compared to pbccs. The models are trained on the data collected from CHM13 cell line (chromosomes 1-18, 2 SMRT Cells), hyper-parameters are optimized using CHM13 cell line (chromosomes 21-22, 2 SMRT Cells), and the models are tested on a completely held-out genome HG002 (chromosome 20, 1 SMRT Cell). 
Here, we use the KL loss for matching teacher and student probabilities and initialize the student model from top layers of the teacher. 
Model runtime is measured using number of seconds on a Google Cloud VM with Intel 32 CPU cores (n1-standard-32).

In order to assess whether distillation is useful, 
we also train student models of the same size but without the distillation loss, only with $L(y, p_{student})$. 
We show that distillation consistently results in improved yield at empirical Q30 over CCS across different student architectures (3, 4, 5 layers) and replicates (Fig.~\ref{fig:runtime_vs_accuracy}b).

Some student models, for example the 5 layer student model, \textbf{achieve a similar yield at empirical Q30 over CCS as the teacher while taking 90\% of teacher's runtime}. 
For the remaining analysis, we focus on the student model with 4 layers, 1.3x faster and 1.5x smaller (in terms of \# parameters) than the teacher, and slight performance degradation (1.69 vs 1.73) compared to the teacher.

\begin{figure}[b]
\begin{floatrow}
\ffigbox{%
  \includegraphics[width=0.35\textwidth]{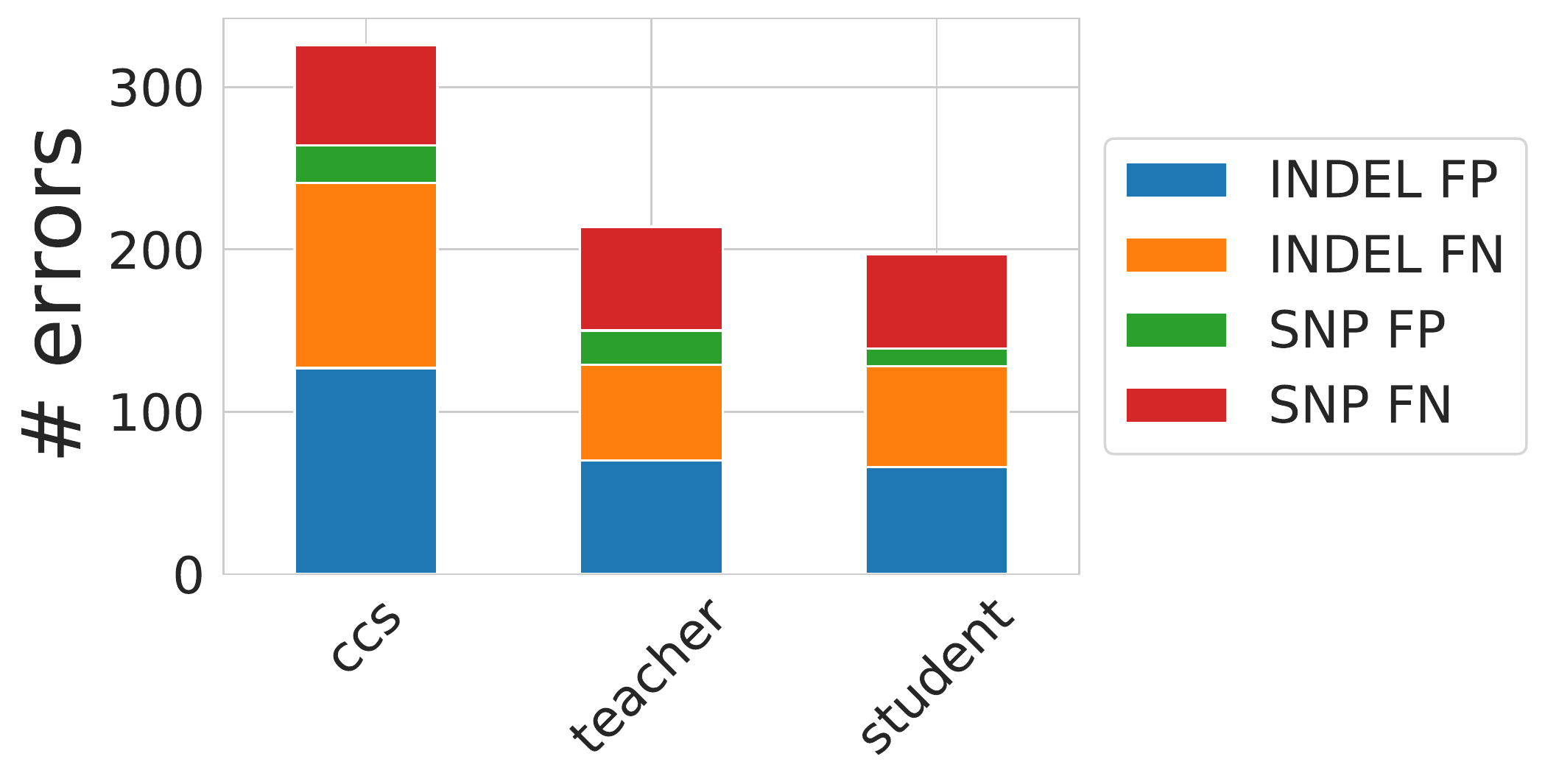}
}{%
  \caption{Distilled DeepConsensus improves variant calling: HG002 (chromosome 20) variant-calling performance of DeepVariant with pbccs, DeepConsensus and Distilled DeepConsensus reads from 2 SMRT Cells. }%
}
\capbtabbox{%
{\fontsize{8}{10}\selectfont
\tabcolsep=0.08cm
\begin{tabular}{
>{\columncolor[HTML]{FFFFFF}}c 
>{\columncolor[HTML]{FFFFFF}}c 
>{\columncolor[HTML]{FFFFFF}}c 
>{\columncolor[HTML]{FFFFFF}}c }
\multicolumn{1}{l}{\cellcolor[HTML]{FFFFFF}{\color[HTML]{000000} Model}} & \multicolumn{1}{l}{\cellcolor[HTML]{FFFFFF}{\color[HTML]{000000} Mean QV ($\uparrow$)}} & \multicolumn{1}{l}{\cellcolor[HTML]{FFFFFF}{\color[HTML]{000000} Complete (\%, $\uparrow$)}} & \multicolumn{1}{l}{\cellcolor[HTML]{FFFFFF}{\color[HTML]{000000} Duplicated (\%, $\downarrow$)}} \\ \hline
{\color[HTML]{000000} pbccs}                                             & {\color[HTML]{000000} 49.8625}                                             & {\color[HTML]{000000} 96.893}                                                    & {\color[HTML]{000000} 0.572}                                                       \\
{\color[HTML]{000000} Teacher}                                        & {\color[HTML]{000000} \textbf{51.999}}                                     & {\color[HTML]{000000} \textbf{97.094}}                                           & {\color[HTML]{000000} \textbf{0.502}}                                              \\
{\color[HTML]{000000} Student}                                        & {\color[HTML]{000000} 51.782}                                              & {\color[HTML]{000000} 97.063}                                                    & {\color[HTML]{000000} 0.516}                                                       \\ \hline
\end{tabular}
\label{table:assembly}
} 

}{%
  \caption{Mean quality, gene completeness, and false duplication rates of the genome assemblies generated with pbccs, DeepConsensus and Distilled DeepConsensus reads from 2 SMRT Cells.}%
}
\end{floatrow}
\end{figure}

\subsection{Teacher versus student error profiles}

We analyze whether the teacher and the student make similar kinds of errors. We quantify the number of base pairs with mismatches, insertions and deletions per read and additionally subdivide errors by their sequence context. It has been shown that PacBio HiFi sequencing has higher error rates within homopolymer (HP) regions~\cite{wenger2019accurate}, which are regions with consecutive repeating bases (e.~g.~AAAA), and thus we split indel errors by HP and non-HP context. Fig.~\ref{fig:runtime_vs_accuracy}c shows that the error types made by the teacher and the student are similar with HP indels and in particular HP insertions for the student being the major error type. 


\subsection{Performance on downstream tasks: variant calling and assembly}


\textbf{Variant calling} We perform variant calling analysis using a custom DeepVariant~\cite{poplin2018deepvariant} model (warm-started from DeepVariant 1.4) trained  on sequences from 2 SMRT Cells generated by DeepConsensus teacher and student models respectively and evaluated on held-out chromosome 20. For sequences output by pbccs, we use DeepVariant 1.4 (trained on pbccs output).
Fig.~2 shows that the student model has 39\% error reduction in variant calling in comparison to pbccs with the student being slightly better than the teacher on this task (34\% reduction for larger model).


\textbf{Assembly analysis} Given increased yield and higher quality reads generated by the DeepConsensus student model and the larger teacher model, we compare the student and the teacher models in terms of \textit{de novo} assembly generation. Phased assemblies were generated with hifiasm assembler using reads (from 2 SMRT Cells) output by pbccs, and DeepConsensus student and teacher models, similar to~\cite{baid2022deepconsensus}. In Table 1, we report the estimated qualities, using YAK~\cite{cheng2021yak}, of the assemblies and show that the mean quality value (across haplotypes) is similar between the teacher and the student models with both being better than pbccs (3.8\% and 4.2\% for student and teacher, respectively). Additionally, both the teacher and the student models increase assembly gene completeness and decrease false gene duplication rate.

  

\begin{figure}[b]
     \centering
     \begin{subfigure}[b]{0.14\textwidth}
         \centering
         \includegraphics[width=\textwidth]{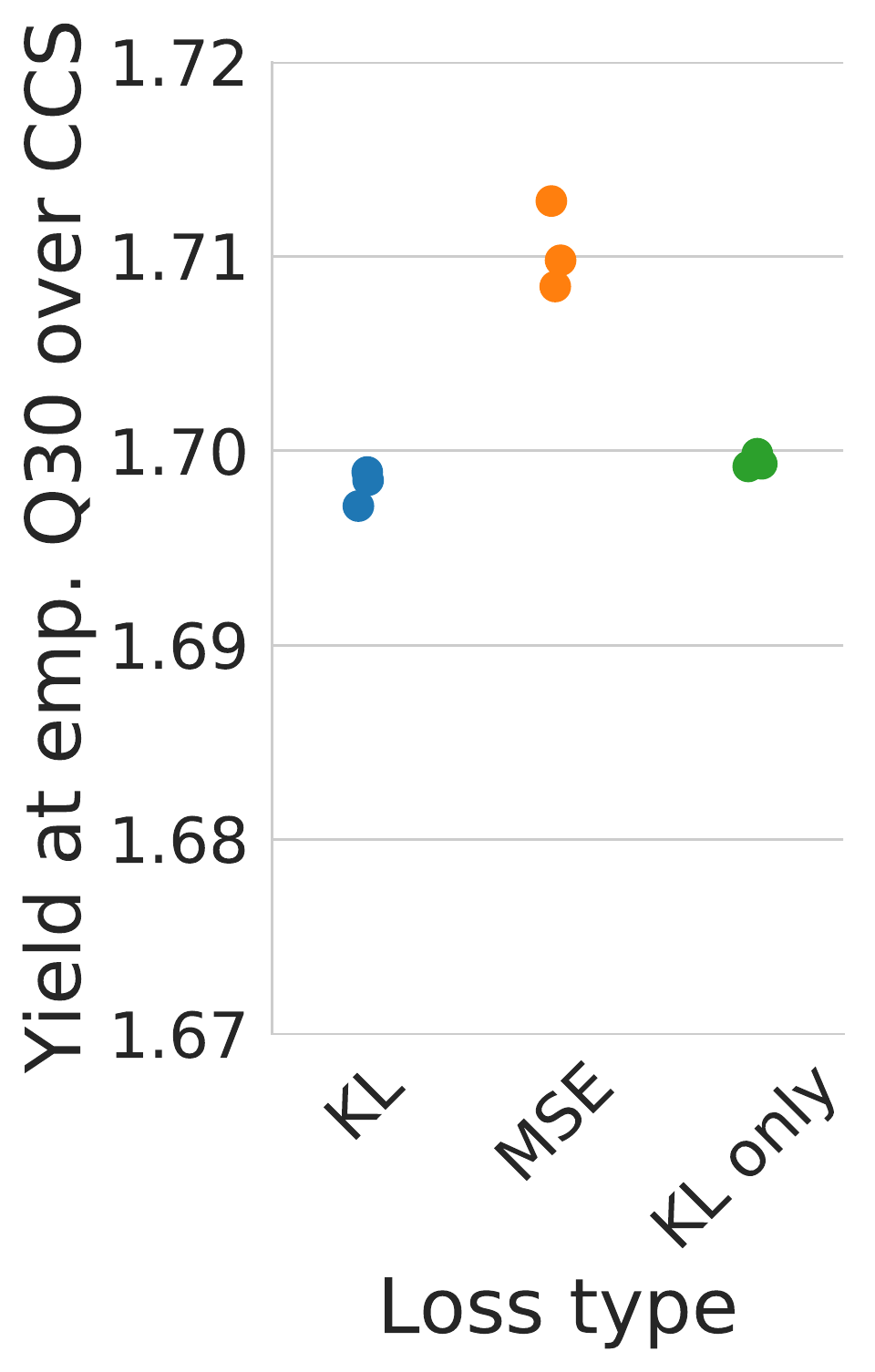}
         \caption{}
     \end{subfigure}
     \hspace{-2em}
     \begin{subfigure}[b]{0.223\textwidth}
         \centering
         \includegraphics[height=\textwidth]{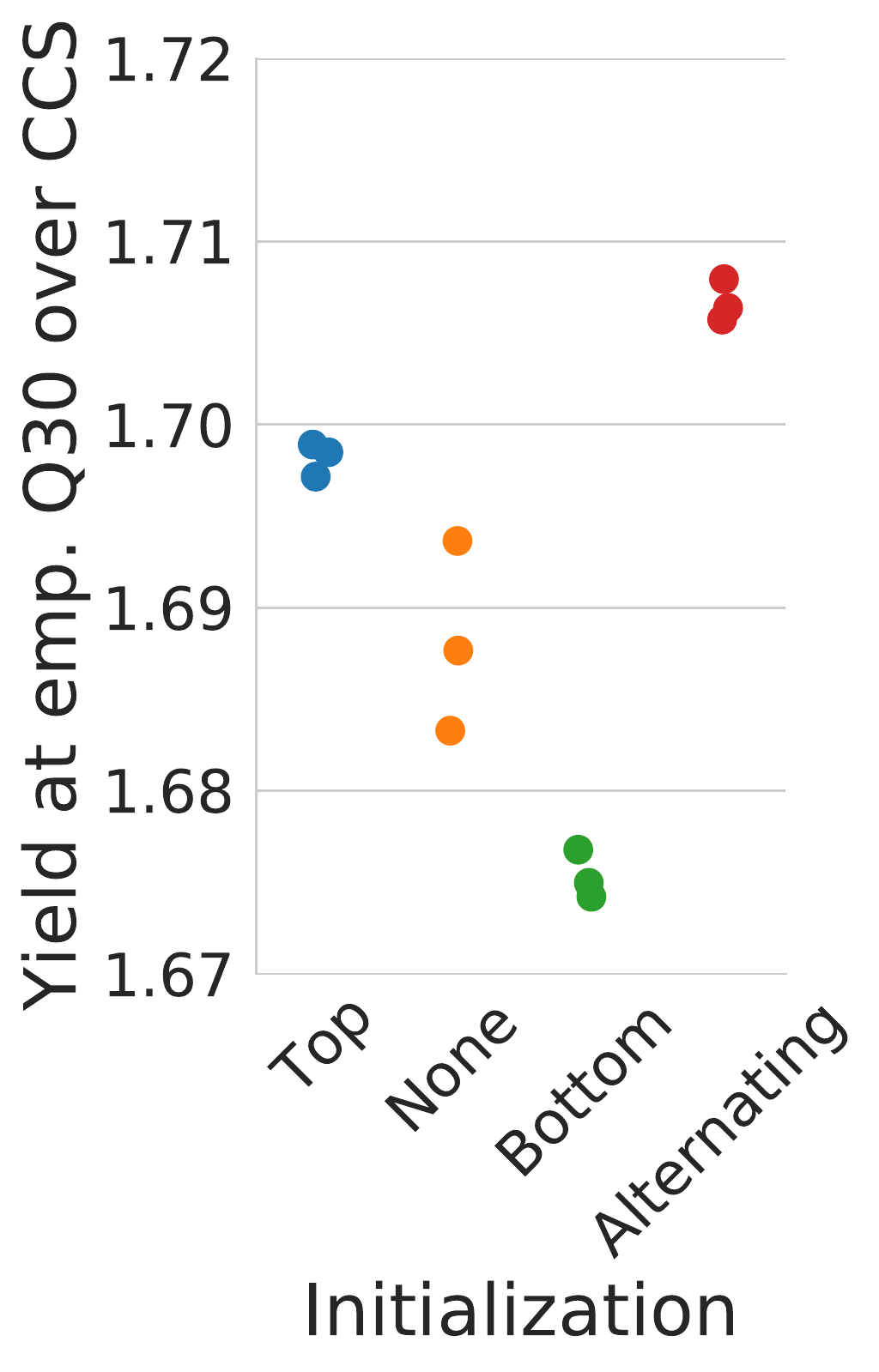}
         \caption{}
     \end{subfigure}
     \hspace{-2em}
     \begin{subfigure}[b]{0.265\textwidth}
         \centering
          \includegraphics[width=\textwidth]{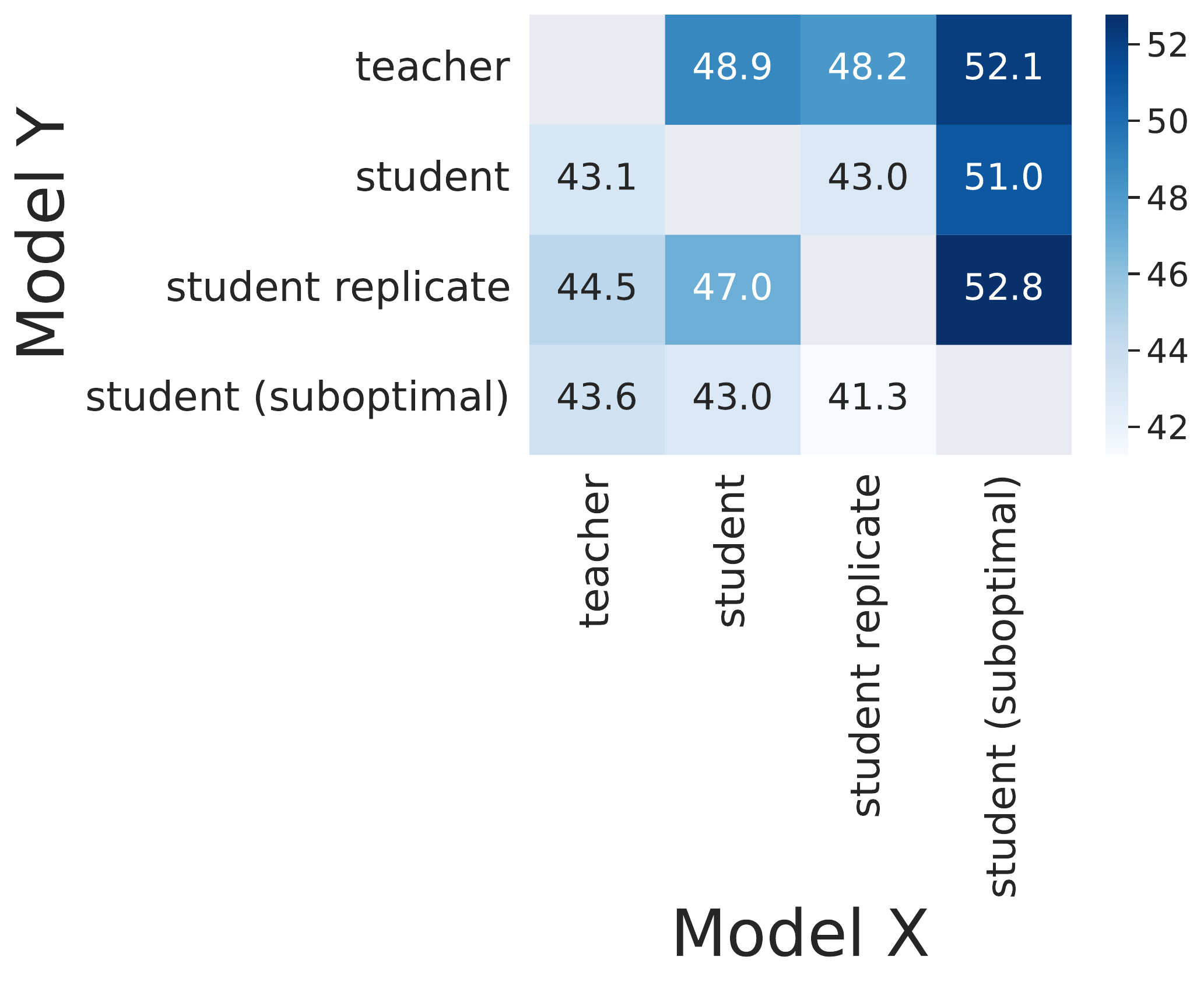}
         \caption{}
     \end{subfigure}
     \begin{subfigure}[b]{0.38\textwidth}
         \centering
          \includegraphics[width=\textwidth]{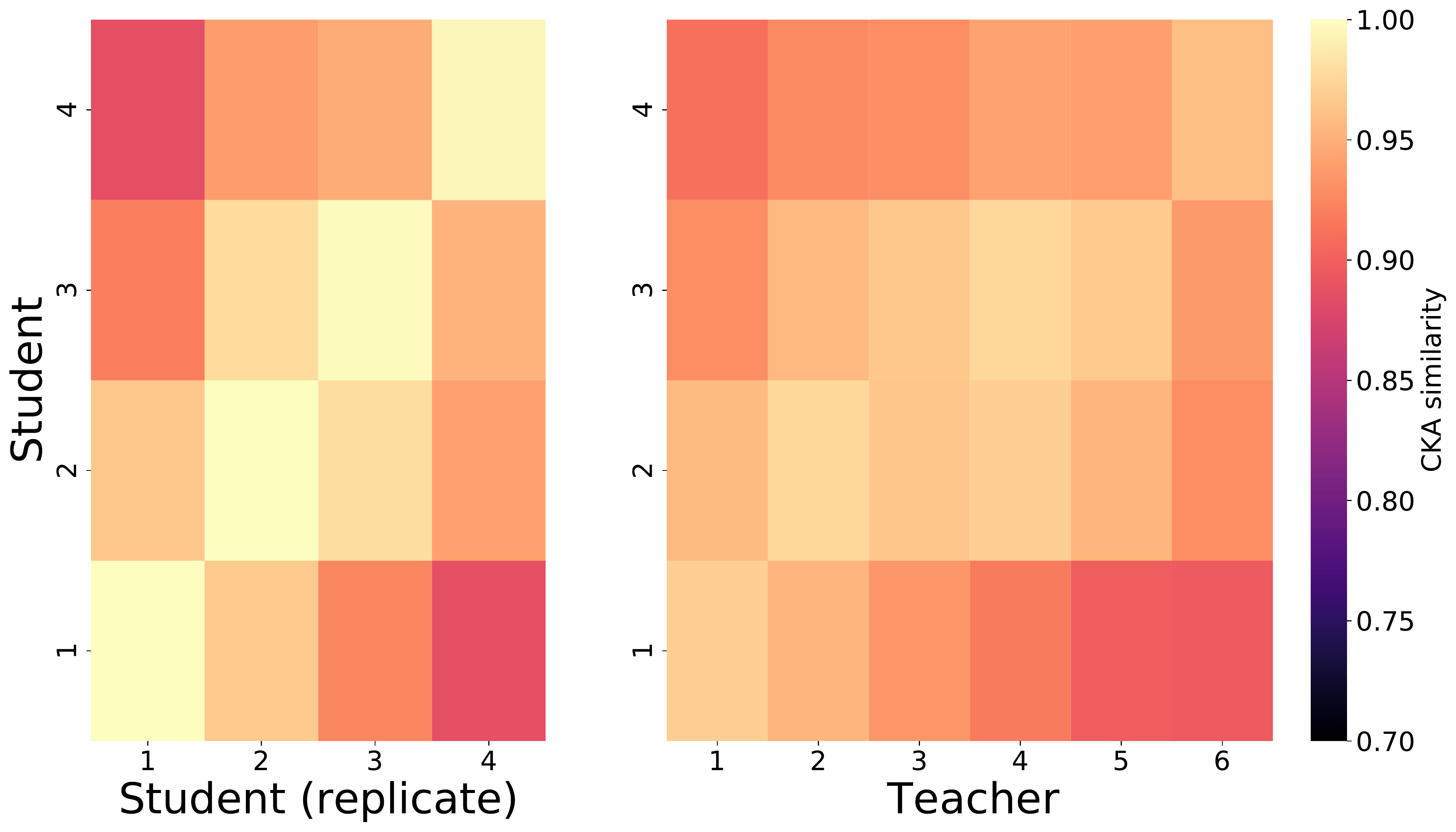}
         \caption{}
     \end{subfigure}
  \caption{
  a) Effect of loss choice on distillation. b) Effect of initialization on distillation. 
  c) Per-example agreement between the teacher, students with the same hyperparameters but different replicates, and students with different less optimal hyperparameters (we take the lowest performing student from Fig.~\ref{fig:analysis}b. We measure the probability that model Y is strictly better (>, equality is omitted and thus probabilities do not add up to 100\%) than model X (in terms of alignment identity) given that model X and Y disagree on the prediction (edit distance > 0). 
  d) CKA analysis between student and a replicate (left), showing representations are stable and CKA between student and teacher (right), showing representations are similar.
  }
  \label{fig:analysis}
\end{figure}

\vspace{-0.7em}

\section{Analysis}


We performed a study of distillation training options to evaluate which factors contributed the most to good performance in the student models. We experimented with three 
distillation losses: alignment to ground truth and Kullback Leibler (KL) divergence with the teacher probabilities, alignment to ground truth and MSE with the teacher probabilities, and just KL with the teacher probabilities.
Additionally, we experimented with four
types of initialization schemes. The relative benefits of each can be seen in Fig.~\ref{fig:analysis}a and b. We found that MSE outperformed KL divergence, and that initializing the student network to "alternating" layers (1, 2, 4, 6 of the teacher) performed best among layer initialization schemes. 

Fig.~\ref{fig:analysis}c shows a per-example agreement analysis~\cite{Shor2022UniversalPS} to better understand whether models with better overall performances are just correctly predicting more examples,
or if they're correctly predicting different 
examples. The results show that the better models are more accurate only slightly more than the worse models when there is a disagreement. This means that the set of examples correctly predicted by the better models is a larger and different set than the set of examples correctly predicted by the worse models, but neither is as a superset of the other.


To test whether teachers and students learn the same representations, we apply linear Centered Kernel Alignment (CKA) between pairs of layers across networks, following the methodology of \cite{raghu2021vision}. CKA computes a [0, 1]-valued similarity between two Gram matrices computed
from two representations over the same sample of input examples~\cite{kornblith2019similarity}. We see from the student vs.~student plot (Fig.~\ref{fig:analysis}d, left) that the internal representations of the student model are very stable across different replicates. 
In the teacher vs.~student plot (Fig.~\ref{fig:analysis}d, right), we find that layers in the student network correspond well with particular layers in the teacher. Interestingly, we find that despite initializing with top layers from the teacher ($1 \rightarrow 1, 2 \rightarrow 2, 4 \rightarrow 3, 6 \rightarrow 4$), the student layers after training have a different correspondence ($1 \rightarrow 1, 2$ or $4 \rightarrow 2, 4 \rightarrow 3, 6 \rightarrow 4$).
This representational difference might also explain why the multiheaded attention of the student is
different than the teacher: for example for layer 3, the student attention head looks an average of 1.8 base pairs different than the teacher for layer 4.











\medskip

{
\small
\bibliographystyle{IEEEtran}
\bibliography{main}
}

\end{document}